\documentclass[review]{elsarticle}
\usepackage{graphicx}
\usepackage{amssymb}
\usepackage{amsmath}

\journal{Journal of Magnetism and Magnetic Materials}

\begin{document}

\begin{frontmatter}

\title{FeCuNbSiB thin films with sub-Oersted coercivity}

\author[1]{J.M. Alves\corref{cor1}}
\ead{juciane@cbpf.br} 
\author[1]{D.E. Gonzalez-Chavez}
\author[1]{N.R. Checca}
\author[1]{B.G. Silva}
\author[1]{R.L. Sommer\corref{cor1}}
\ead{sommer@cbpf.br} 

\cortext[cor1]{Corresponding authors}
\affiliation[1]{organization={Centro Brasileiro de Pesquisas Físicas},
                postcode={22290-180},
                city={Rio de Janeiro},
                country={Brazil}}

\begin{abstract}
Nanocrystalline FeCuNbSiB thin films were fabricated through magnetron sputtering followed by heat treatment, resulting in samples characterized by low coercivity and high effective magnetization. Comprehensive microstructural analysis, employing X-ray diffraction and transmission electron microscopy techniques such as selected area electron diffraction, high-resolution imaging, and Fourier transform, was conducted. Magnetic properties were investigated using an alternating gradient field magnetometer and broadband ferromagnetic resonance.
The structural analysis revealed a well-defined microstructure of nanograins within an amorphous matrix in all of our films. However, the coercivity of the 80 nm films did not exhibit as low values as observed for the 160 nm films

\end{abstract}


\end{frontmatter}
\section{Introduction}

Nanocrystalline magnetic materials have received great attention due to their soft magnetic properties \cite{Yoshizawa1988, Yoshizawa1991, Herzer1992a, Herzer1992b, Herzer1997, Herzer2014} and possibility of several important applications in magnetic devices, such as magnetic sensors, transformer, materials for shielding, and magnetic integrated circuits. 
Among nanocrystalline magnetic materials, the FeSiB-based magnetic alloy stands out, specially the so-called FINEMET alloy with typical composition $\mathrm{Fe_{73.5}Si_{13.5}B_{9}Nb_{3}Cu_{1}}$. 
The soft magnetic properties (low coercive field, high permeability and high saturation magnetization) are in general associated with the nanocrystalline state and the corresponding quenching of magnetocrystalline anisotropy \cite{Yoshizawa1988, Yoshizawa1991, Herzer1993, Herzer1995, McHenry1999}.
These properties have been primarily achieved with the material in the form of a ribbon, and the process is well-established in the literature, resulting in optimized magnetic properties. These optimized properties include a saturation magnetization as high as 987 $\mathrm{emu/cm^3}$ and coercive field as low as 0.007 Oe \cite{Yoshizawa1988}.
The nanocrystalline microstructure of bulk amorphous magnetic alloys in ribbon form is characterized by the partial crystallization of FeSi phase ($\mathrm{DO_{3}}$ structure) \cite{Ayers1997, Hono1998, McHenry1999} with crystallite sizes in the range of 10 nm to 15 nm. These crystallites are randomly distributed within a residual amorphous matrix.
In these grains, which are smaller than the exchange length, exchange interactions force the magnetic moments to align in parallel, presenting a challenge for magnetization to conform to the preferred local anisotropies inside the grains.
In this way, effective magnetocrystalline anisotropy is the average over several randomly oriented grains \cite{Herzer1997, Herzer1993, Herzer1995, Hono2000}, resulting in a reduced effective anisotropy amplitude. 
This phenomenon also contributes to reducing the coercivity of the material. 
Coercivity is influenced not only by magnetocrystalline anisotropy but also by the magnetostriction.
The latter is diminished  through the development of negative magnetostriction in the ferromagnetic phase immersed in the amorphous matrix with positive magnetostriction \cite{Herzer1992b, Herzer1997}.
To achieve the desired magnetic properties, all these processes must be carefully controlled through the chemical composition of the bulk alloy and the heat treatment process.
The copper added to the alloy forms clusters with diameters of 2 nm to 5 nm at a temperature approximately 50 °C below the crystallization onset temperature.
These copper clusters function as nucleation sites for FeSi, and in an optimized microstructure, they are surrounded by FeSi nanocrystals \cite{Ayers1997, Herzer2014, Gheiratmand2016}.
The residual amorphous matrix is stabilized by the presence of boron and niobium which have low solubility in FeSi \cite{Yoshizawa1991, Herzer1997}. This results in a limiting process to the grain size growth.
Kinetic and thermodynamic studies indicate that the annealing of the unstable amorphous phase results in the formation of a metastable $\mathrm{Fe_{3}Si}$ crystalline phase with an ordered $\mathrm{DO_{3}}$ structure \cite{Herzer1992a, Herzer1992b, Hono1998, Hono2000}. 

To integrate these materials into micro and nanofabrication techniques, it becomes imperative to replicate the same magnetic properties at a smaller scale, specifically, in the form of nanometric thin films where other sources of magnetic anysotropy are likely to appear. To control the overall soft magnetic properties in thin films of these materials becomes then a challenge that makes difficult to obtain sub-Oersted coercive fields. The possibility miniaturization of devices using nanocrystalline magnetic films of high permeability and high saturation values drives continuously the efforts to tailor the soft magnetic properties of these materials an nanometric thickness.

In this study we produced FeCuNbSiB nanocrystalline thin films (80 nm and 160 nm) through a heat treatment process (490 ºC, 500 ºC and 520 ºC), resulting in samples with sub-Oe coercivity and high effective magnetization.
The structural and magnetic properties of these films were analyzed by means of X-ray diffraction, transmission electron microscopy, alternating gradient field magnetometry and broadband ferromagnetic resonance.

\section{Experimental}

Thin films of 80 nm and 160 nm thicknesses were prepared by magnetron sputtering from a target with nominal composition of $\mathrm{Fe_{73.5}Si_{13.5}B_{9}Nb_{3}Cu_{1}}$, produced by ACI Alloys INC, and deposited onto a $\mathrm{Si(100)/SiO_2}$ substrate with a 300 nm oxide layer. 
The depositions were carried out at room temperature at a rate of 1.29 Å/s, using a DC source with 50 W of power, working pressure of 5 mTorr and Ar gas flow of 50 sccm after a base pressure of $1\times 10^{-8}$ Torr was reached in the sputtering chamber.
The crystallization induction occurred through vacuum annealing ($1\times 10^{-5}$ Torr) for 1 hour at temperatures of 490 °C, 500 °C and 520 °C, employing heating and cooling rates of 5 °C/min.

The structural properties of the films were investigated via grazing incidence X-ray diffractometry (GIXRD) with a Panalytical X'PERT PRO MRD diffractometer with Cu-K$\alpha$ radiation ($\lambda$=1.54056 Å). 
The diffraction peaks were indexed to identify the crystallographic phases.
A pseudo-Voigt profile function was employed to estimate the line broadening parameter $\beta$ at half the maximum intensity (FWHM) for the $\mathrm{Fe_3Si}$ (220) peak around the Brag angle $2\theta\approx$ 45°. 
Subsequently, the Scherrer equation (Eq. \ref{eq:Scherrer}) was used to determine the crystallite size $D$.

\begin{equation}
    \beta = \frac{0.9 \lambda}{D \, \cos \theta}
    \label{eq:Scherrer}
\end{equation}

A microstructural investigation using transmission electron microscopy (TEM) was conducted, involving the preparation of lamellas using focused ion beam (FIB) with a gallium ion source from the FIB LYRA-3 double beam microscope (TESCAN). 
To protect the film, a platinum layer was deposited in situ using a gas injection system, followed by sequential cuts by sputtering layer-by-layer with the Ga ion source. 
The lamellas obtained were then fixed on a copper TEM grid.
A thinning process was initiated by applying a probe current of 1 nA and completed with a current of 50 pA.
FIB-prepared lamellas were analyzed by a JEOL 2100F transmission electron microscope operating at 200 kV and coupled with a CMOS (ONE VIEW) to acquire TEM and high resolution TEM (HRTEM) images, as well as selected area electron diffraction (SAED) patterns. 

Magnetic properties were examined through in-plane (M vs H) magnetization curves obtained with an alternating gradient field magnetometer (AGFM) and through broadband ferromagnetic resonance (FMR) using a vector network analyzer based setup.
For both measurements an in-plane magnetic field with a maximum amplitude of ± 600 Oe was applied parallel to the sample plane.
Effective magnetization $M_\mathrm{eff}$ and in-plane anisotropy $H_{k}$ parameters were derived from broadband-FMR spectra by fitting the measured dispersion relation. 
This fitting process involved employing the Kittel equations, aligning resonance frequency $f_{r}$ with applied field  $H$ curves.
For the easy axis 0° Eq. \ref{eq:kittel0} was used, while for hard axis 90° Eq. \ref{eq:kittel90} was employed.
\begin{equation}
   f_{r}=\frac{\gamma}{2\pi}
   \sqrt{(4\pi M_\mathrm{eff} \pm H + H_k)(\pm H + H_k)}
\label{eq:kittel0}
\end{equation}

\begin{equation}
\centering
   f_{r}=\frac{\gamma}{2\pi}
   \sqrt{(4\pi M_\mathrm{eff} \pm H)(\pm H - H_k)}
\label{eq:kittel90}
\end{equation}

The effective magnetization $M_\mathrm{eff}$ is related to the saturation magnetization $M_s$ and the out-of-plane anisotropy field $H_\perp$ present in the sample ($4\pi M_\mathrm{eff} = 4\pi M_s - H_\perp$).
Obtaing $M_\mathrm{eff}$ form the broadband FMR is a straightforward process as it does not depend on the total volume of the measured sample. 
In contrast, estimating $M_s$ form traditional magnetometry requires a accurate measurement of sample thickness and area, along with magnetometer calibration for the small magnetic moments of thin film samples.

\section{Results and discussion}

Figure 1 shows the diffraction patterns for the 80 nm and 160 nm films. 
It was observed that the as-deposited films exhibited a broad peak centered at 44.52° and 44.25° for 80 nm and 160 nm, respectively.
These peaks are characteristic of amorphous alloys with short-range ordering \cite{McHenry1999}.
After heat treatment, the position of the major peak, corresponding to the $\mathrm{Fe_3Si}$ (220) phases showed up close the the amorphous matrix peak.
However, a peak displacement attributable to microstructural changes \cite{McHenry1999} is noticeable.
The average peak displacement after heat treatment are 1.58° and 1.92° for the 80 nm and 160 nm films, respectively.

\begin{figure*}[h!]
\centering
\includegraphics[width=15cm]{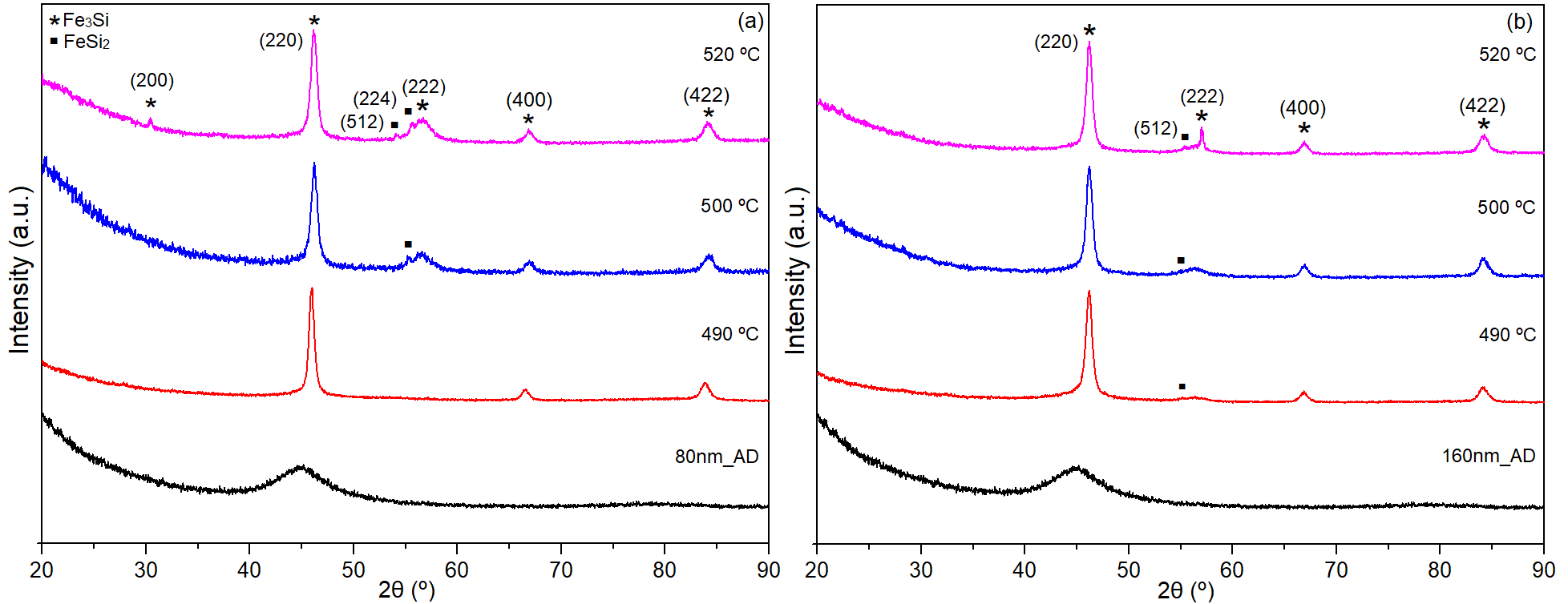}
\caption{Diffraction patterns of FeCuNbSiB films with 80 nm (a) and 160 nm (b), as-deposited $AD$, and after annealing at 490 °C, 500 °C and 520 °C. The symbols represent the indexed diffraction planes for $\mathrm{Fe_3Si}$ (stars) and $\mathrm{FeSi_2}$ (squares).}
\label{fig:RX}
\end{figure*}

In both films, the crystallization of the $\mathrm{Fe_{3}Si}$ phase is predominant. 
This was evidenced by the XRD patterns displaying diffraction peaks corresponding to (220), (400), and (422) planes, considering an ordered $\mathrm{DO_{3}}$ structure, for all annealing temperatures.
The $\mathrm{DO_{3}}$ structure was confirmed by TEM SAED measurements, which also showed sets of equivalent lattice planes $\{220\}$.
Nevertheless, small peaks of the non-ferromagnetic $\mathrm{FeSi_{2}}$ phase, with an orthorhombic structure, were observed at different annealing temperatures.

The mean $\mathrm{Fe_{3}Si}$ lattice parameter ($a_m$) estimated by XRD, as well as the particle size ($D$), are presented in Table 1. 
The results showed small variations in the lattice parameter for both samples as a function of annealing temperatures and film thickness.
These variations arise from slight fluctuations in the atomic percentage of silicon content dissolved in the iron unit cell of the $\mathrm{Fe_{3}Si}$ grains \cite{Rixecker1993, Gheiratmand2016}, influenced by variables such as the chemical composition of the alloy, annealing temperature and nucleation rate.
An increase in the silicon atomic percentage in the FeSi crystal structure results in a reduction in the lattice parameter.
Our results are compatible with Si atomic percentages above 19 \% \cite{Zemcik1991} in the crystallized grains, which are expected to show an ordered $\mathrm{DO_{3}}$ structure instead of the disordered BCC structure observed for lower Si atomic percentages \cite{Gheiratmand2016, Rixecker1993}.
Additionally, higher Si atomic percentages reduce the grain anisotropy \cite{Herzer1997}, which also reduces the overall alloy anisotropy.

The crystallite size $D$, estimated by XRD and corroborated by TEM (Table \ref{tab:Results}), fell within the expected range for FINEMET, between 10-15 nm \cite{Yoshizawa1988, Herzer1993, Gheiratmand2016}.
These sizes remained nearly constant at different annealing temperatures for both 80 nm and 160 nm films.
Only for the 80 nm  film at 490 °C, a slightly larger grain size was observed.

\begin{table*}[h!]
\caption{Lattice parameters and particle sizes of FeCuNbSiB films after annealing at 490 °C, 500 °C and 520 °C.}
 \centering
  \begin{tabular}{ c c c c c c}
   \hline 
    Sample & $T_{ann.}$ (°C) & $\mathrm{Fe_{3}Si}$ $a_m (Å)$  & $D$(nm) XRD & $D$(nm) TEM\\
     
   \hline 
          & 490 & 5.6157 & 14.1 & 15.5 \\
    80 nm & 500 & 5.6060 & 11.9 & 12.0 \\
          & 520 & 5.6588 & 12.2 & 11.9 \\
   \hline 
           & 490 & 5.6128 & 12.2 & 11.6 \\
   160 nm  & 500 & 5.6114 & 12.2 & 11.1 \\
           & 520 & 5.5922 & 12.7 &  -- \\
\hline
\end{tabular}
\label{tab:Results}
\end{table*}

TEM results are summarized in Fig. 2 and Fig. 3.
These figures show high resolution images of cross sections of the films, including an enlarged image of the as-deposited films. They also display SAED patterns of the circled area of the film and statistics of grain size counts for the nanocrystallized films. 

\begin{figure*}[h!]
\centering
\includegraphics[width=10cm]{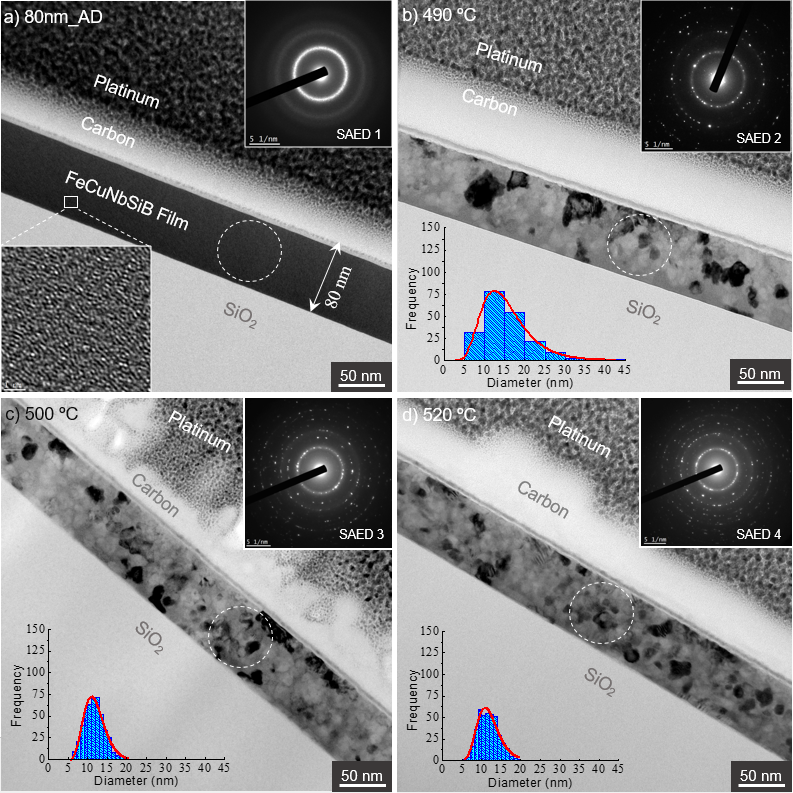}
\caption{Selected-area electron diffraction and TEM images of FeCuNbSiB thin films with 80 nm, $AD$ (a) and after annealing at (b) 490 °C, (c) 500 °C and (d) 520 °C.}
\label{fig:TEM_80nm}
\end{figure*}

\begin{figure*}[h!]
\centering
\includegraphics[width=10cm]{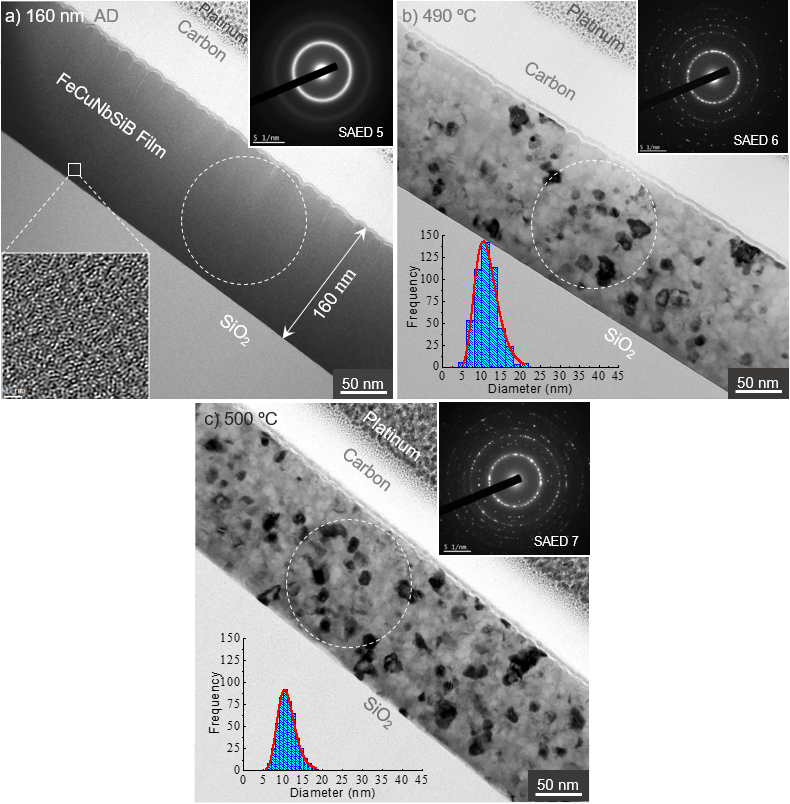}
\caption{Selected-area electron diffraction and TEM images of FeCuNbSiB thin films with 160 nm, $AD$ (a) and after annealing at (b) 490 °C and (c) 500 °C.}
\label{fig:TEM_160nm} 
\end{figure*}

The SAED patterns for the as-deposited films evidence a short-range ordering, characteristic of the amorphous alloys \cite{Herzer2014, Hono1998, Hono1992, Hono1993}.
On the other hand, annealed films show SAED patterns typical of nanocrystalline samples \cite{Herzer1993, Hono2000}. 
The observed SAED patterns are coherent with the XRD measurements, indicating the predominance of the $\mathrm{Fe_{3}Si}$ phase.

The particle size distribution (PSD) were obtained by analyzing various TEM images for each sample. 
This distribution is highly asymmetric.
The mean grain $D$ size value was obtained by fitting the PSD to a lognormal distribution. These values are presented in Table \ref{tab:Results}.

TEM analysis revealed the absence of Cu clusters, and the literature suggests that in the optimized microstructure, Cu clusters are typically enveloped by FeSi nanocrystals \cite{Hono1998}.

\begin{figure*}[h!]
\centering
\includegraphics[width=5cm]{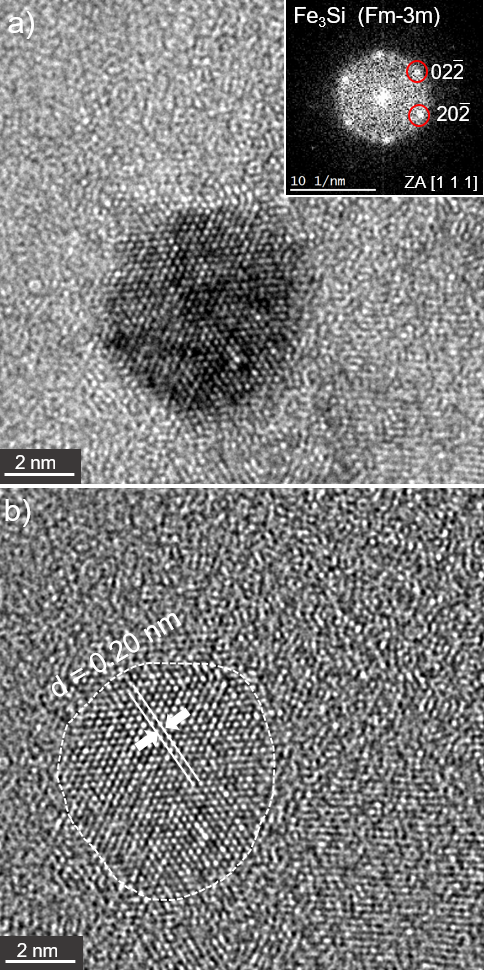}
\caption{Fourier transform (a) and interplanar distance (b) of FeCuNbSiB thin film with 160 nm after annealing at 500 °C.}
\label{fig:FFT_TEM}
\end{figure*}

Figure \ref{fig:FFT_TEM} shows fast Fourier transform (FFT) of one crystallized grain and HRTEM image of the selected areas within and outside the grain boundaries.
The internal region exhibited a clear crystalline pattern, and fast Fourier transform indexing analysis revealed $\mathrm{Fe_{3}Si}$ phase (Fig. 4a). 
Direct measurements of interplanar distances yielded a value of 0.20826 nm, consistent with this identified phase (Fig. 4b).
Furthermore, the external HRTEM image clearly indicates that this area corresponds to the residual amorphous matrix.
Performing the same analysis on other portions of our samples yielded consisting results. However, it is worth noting that a minimal number of $\mathrm{FeSi_{2}}$ grains were observed, though they are not presented here.
 
\label{sec:magnetic resutls}

\begin{figure*}[h!]
\centering
\includegraphics[width=6cm]{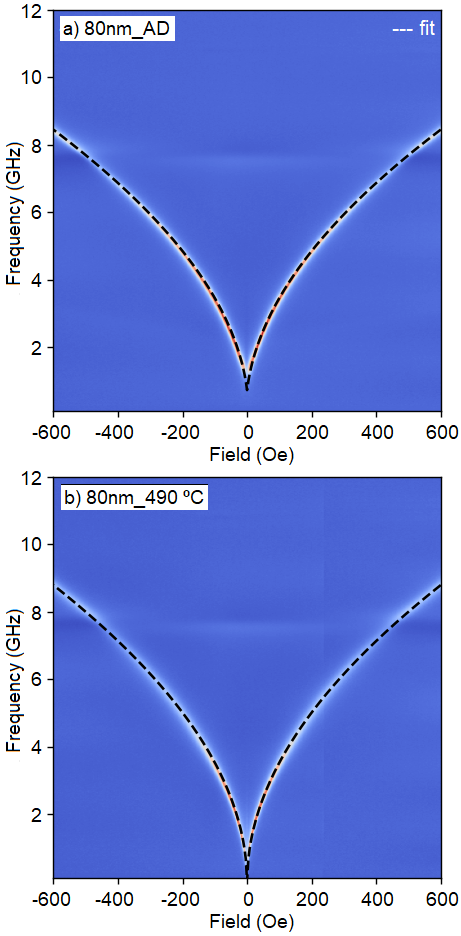}
\caption{Representative broadband-FMR spectra (two-dimensional color plots) and dispersion relation fits (dashed lines) for FeCuNbSiB thin film with 80 nm thickness.}
\label{fig:VNA_FMR}
\end{figure*}

Figure \ref{fig:VNA_FMR} illustrates a typical broadband-FMR spectrum. 
The color scale represents the FMR absorption amplitude. 
Resonant peaks, where FMR absorption is maximal, are represented as branches following a V-shaped dispersion relation (resonant frequency vs. resonant field), characteristic of soft magnetic materials.
Branch inclination correlates with the effective magnetization, while anisotropies $H_k$ induce frequency shifts in the dispersion relations. 
Precise values for these parameters are obtained by fitting the dispersion relation to the Kittel equations (Eq. \ref{eq:kittel0} and Eq. \ref{eq:kittel90}).

In addition to analyzing the effective magnetization ($M_\mathrm{eff}$) and effective anisotropy ($H_k$) obtained from broadband-FMR measurements, the coercivity ($H_c$) was also examined from magnetization curves.
The obtained values for these parameters, depending on the sample thickness and heat treatment temperature, are detailed in Table \ref{tab:Results2}.

\begin{table*}[h!]
\caption{Coercive field, effective magnetization and effective anisotropy of FeCuNbSiB thin films with 80 and 160 nm after annealing at temperatures of 490 °C, 500 °C and 520 °C.}
 \centering
  \begin{tabular}{ c c c c c }
   \hline 
    Sample & $T_{ann.}$ (°C) & $H_{c}$ (Oe) & $M_{eff}$ ($\mathrm{emu/cm^3}$) &  $H_{k}$ (Oe)\\
     \hline  
          & As-deposited & 3.13 & 1170 & 10 \\
          & 490 & 1.79 & 1290 & 0 \\
   80 nm  & 500 & 1.13 & 1280 & 0 \\
          & 520 & 1.29 & 1280 & 0 \\
   \hline 
           & As-deposited & 2.45 & 1220 & 2 \\
           & 490 & 0.51 & 1260 & 0 \\
   160 nm  & 500 & 0.62 & 1260 & 0 \\
           & 520 & 0.63 & 1250 & 0 \\
    \hline
  \end{tabular}
\label{tab:Results2}
\end{table*}

As expected, heat treatment at these temperatures resulted in a substantial reduction in the coercive field for all treated samples compared to the as-deposited ones.
The 160 nm sample, treated at 490 °C, exhibited the lowest observed coercive field (0.51 Oe). 
Furthermore, all 160 nm treated samples showed sub-Oe coercivities.
On the other hand, for 80 nm samples the lowest coercive field was 1.13 Oe for 500 °C heat treatment.
Heat treatment also resulted in the elimination of uniaxial anisotropy and an increase in effective magnetization. 
This later effect is expected due the effect of the nanocrystallization process and the corresponding quenching of magnetocrystalline anisotropy. 
The obtained coercivity values are still far form the ultra-low coercivities (0.007 Oe) obtained in bulk ribbons, but seem to be good progress when compared to typical values reported in the literature, particularly for 160nm thick films \cite{Sommer1995, Petó1995, Koszó1996, Pászti1998, Neuweiler1998, Li2004, Sommer2006, Celegato2011, Masood2017a, Masood2017b, Mikhalitsyna2020}.

\begin{figure*}[h!]
\centering
\includegraphics[width=13cm]{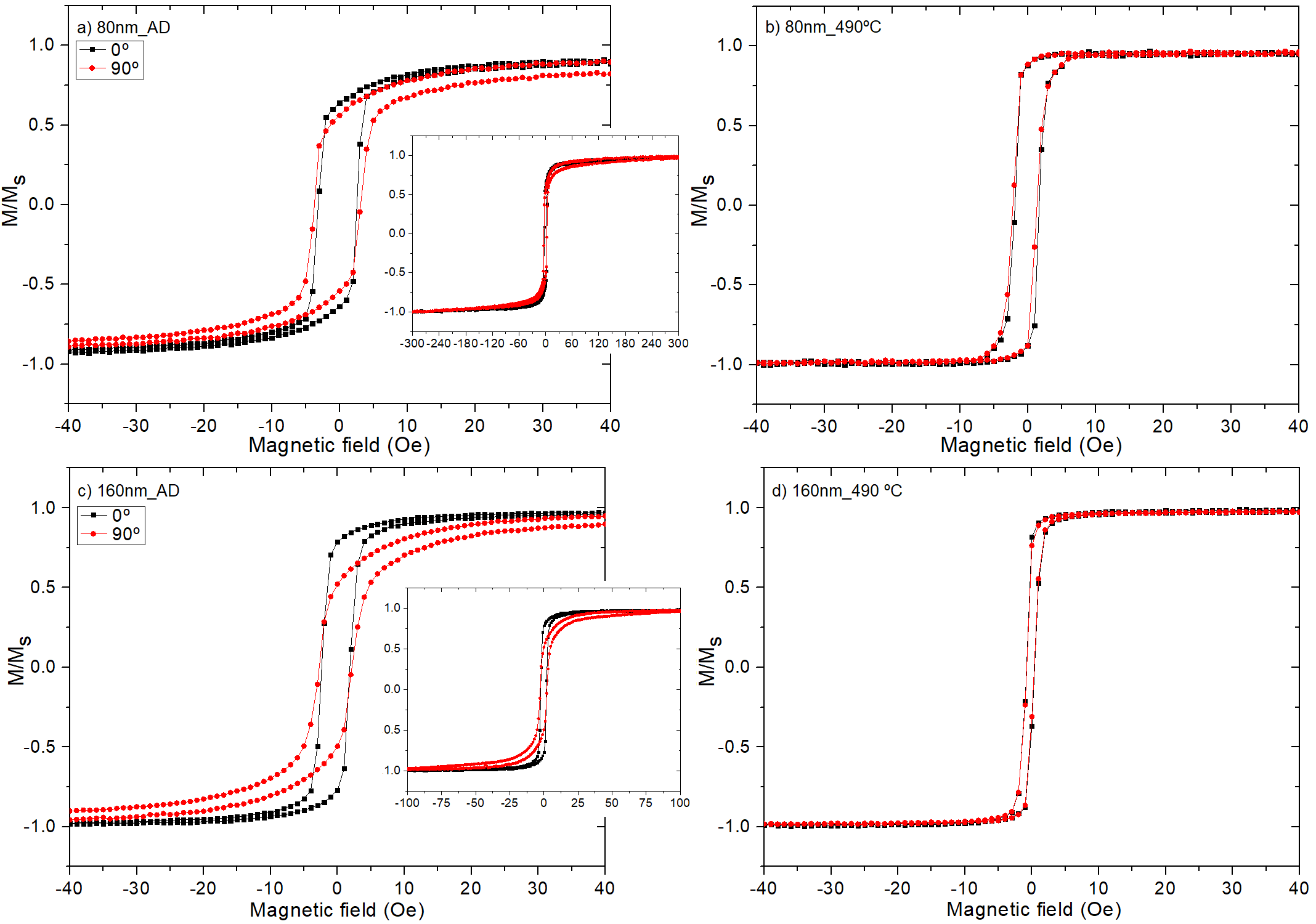}
\caption{Magnetization curves of (a) 80 nm film as-deposited and (b) 80 nm film after 490 ºC heat treatment; (c) 160 nm film as-deposited and (d) 160 nm film after 490 ºC heat treatment. The curves where measured in-plane at two perpendicular angles (0º and 90º).}
\label{fig:MxH}
\end{figure*}

Selected magnetization curves are presented in Fig. \ref{fig:MxH}.
For the as-deposited samples, a small in-plane anisotropy is appreciable, evident from the difference between the 0° and 90° magnetization curves. 
The non-zero effective anisotropy parameter, calculated from FMR dispersion relation fittings, further confirms this observations.
This anisotropy is likely induced during sample growth, and possibly attributed to the magnetic field generated by the sputtering gun.
Additionally, the magnetization curves display rounded profiles, suggesting a probable presence of out-of-plane anisotropy.

The magnetic parameters of as-deposited films exhibit a notable disparity between 80 nm and 160 nm samples, primarily attributable to variations in film thickness.
In particular, the lower effective magnetization for 80 nm can be correlated to the higher out-of-plane anisotropy seen in the magnetization curves. 
This aligns with expectations, as out-of-plane anisotropy is commonly tied to surface interactions and surface roughness effects, showing dependence on film thickness.

The effective magnetization obtained by broadband FMR in the studied samples is larger than the saturation magnetization reported in the literature. 
Yoshizawa \cite{Yoshizawa1988} obtained from FINEMET ribbons a saturation magnetization of approximately 987 $\mathrm{emu/cm^3}$. In turn, Herzer \cite{Herzer1997} attained a saturation magnetization of approximately 1010 $\mathrm{emu/cm^3}$ after heat treatment, while the as-quenched sample exhibited a saturation magnetization close to 971 $\mathrm{emu/cm^3}$.

\section*{Conclusions}

In summary, in this work, we were able to produce nanocrystallized FeCuNbSiB films with a very good microstructure of nanograins in an amorphous matrix containing nanocrystallites of $\mathrm{Fe_{3}Si}$ with sizes between 11 nm and 15 nm, as observed by TEM and XRD. No Cu clusters were observed by TEM.
Also, there was a variation in the lattice parameter of the $\mathrm{Fe_{3}Si}$ phase, possibly due to small fluctuations in the atomic percentage of dissolved silicon in the unit cell.
Interestingly, sub-Oe coercivity was observed for the thicker films, 160 nm, but not for 80 nm. These results may indicate a relationship between magnetostriction and film thickness or effects of surface interactions and surface roughness.
A residual $\mathrm{FeSi_{2}}$ phase observed in XRD and TEM analyses did not influence the magnetic properties of the samples, given its residual content.

From the results obtained in this study with thin films of FeCuNbSiB, it is clear that the crystallization of the $\mathrm{Fe_{3}Si}$ phase led to an increase in effective magnetization after the thermal treatments at the temperatures employed, indicating a significant microstructural change. 
The effective magnetization measured by broadband FMR resulted in values larger than the ones reported in the literature. 
This may be an indication that our sputtering target have more iron than expected, or some kind of strange anysotropy was developed in our sample with the synthesis and annealing processes. 
EDS studies performed on the sputtering targets and produced films have not indicated any extra iron content. 
This is a point to be further addressed in our work.

\section*{Acknowledgment}
 
This wok received support from Brazilian agencies CNPq and FAPERJ. We also acknowledge the LABNANO/CBPF and LABSURF/CBPF for the fabrication infrastructure.

\bibliographystyle{ieeetr}
\bibliography{Refs}

\end
{document}